\documentclass[sigconf]{acmart}
\usepackage{pifont}

\usepackage{stfloats}

\usepackage{hyperref}

\newcommand{\dataset}{MADE-WIC}

\usepackage{xcolor}

\definecolor{red_approach}{HTML}{990000}
\definecolor{green_approach}{HTML}{009900}

\usepackage{listings}
\usepackage{multirow}
\usepackage{colortbl}

\definecolor{codegreen}{rgb}{0,0.6,0}
\definecolor{codegray}{rgb}{0.5,0.5,0.5}
\definecolor{codepurple}{rgb}{0.58,0,0.82}
\definecolor{backcolour}{rgb}{0.95,0.95,0.92}
\lstdefinestyle{code_style}{
    commentstyle=\color{codegray},
    keywordstyle=\color{blue},
    numberstyle=\footnotesize,
    stringstyle=\color{codegreen},
    basicstyle=\ttfamily\footnotesize,
    breakatwhitespace=false,
    breaklines=true,
    captionpos=t, 
    keepspaces=true,
    numbers=left,
    numbersep=5pt,
    showspaces=false,
    showstringspaces=false,
    showtabs=false,
    tabsize=2,
    xleftmargin=5.0ex,
    linewidth=\columnwidth,
    language=c,
    escapeinside={(*@}{@*)}
}
\usepackage{multirow}
\usepackage{makecell}

\setcopyright{acmcopyright}
\copyrightyear{2018}
\acmYear{2018}
\acmDOI{XXXXXXX.XXXXXXX}

\acmConference[ASE 2024]{39th IEEE/ACM International Conference on Automated Software Engineering}{October 2024}{Sacramento, California, United States}
\acmPrice{15.00}
\acmISBN{978-1-4503-XXXX-X/18/06}

\begin{document}

\title{\dataset: Multiple Annotated Datasets for Exploring Weaknesses In Code}

\newcommand{\etal}{\textit{et al.}}

\author{Moritz Mock}
\orcid{0009-0009-3156-6211}
\affiliation{%
  \institution{Free University of Bozen-Bolzano}    \city{Bozen-Bolzano}
  \country{Italy}}
\email{moritz.mock@student.unibz.it}

\author{Jorge Melegati}
\orcid{0000-0003-1303-4173}
\affiliation{%
  \institution{Free University of Bozen-Bolzano}    \city{Bozen-Bolzano}
  \country{Italy}}
\email{jorge.melegati@unibz.it}

\author{Max Kretschmann}
\affiliation{%
  \institution{Hamburg University of Technology}  \city{Hamburg}
  \country{Germany}}
\email{max.kretschmann@tuhh.de}

\author{Nicol\'{a}s E. D\'{i}az Ferreyra}
\orcid{0000-0001-6304-771X}
\affiliation{%
  \institution{Hamburg University of Technology}  \city{Hamburg}
  \country{Germany}}
\email{nicolas.diaz-ferreyra@tuhh.de}

\author{Barbara Russo}
\orcid{0000-0003-3737-9264}
\affiliation{%
  \institution{Free University of Bozen-Bolzano}  \city{Bozen-Bolzano}
  \country{Italy}}
\email{barbara.russo@unibz.it}

\begin{abstract}
In this paper, we present \dataset, a large dataset of functions and their comments with multiple annotations for technical debt and code weaknesses leveraging different state-of-the-art approaches.  
It contains about 860K code functions and more than 2.7M related comments from 12 open-source projects. 
To the best of our knowledge, no such dataset is publicly available. 
\dataset~aims to provide researchers with a curated dataset on which to test and compare tools designed for the detection of code weaknesses and technical debt. 
As we have fused existing datasets,  researchers have the possibility to evaluate the performance of their tools by also controlling the bias related to the annotation definition and dataset construction.
The demonstration video can be retrieved at \url{https://www.youtube.com/watch?v=GaQodPrcb6E}.
\end{abstract}



\maketitle

\section{Introduction}
\begin{figure}[!t]
    \centering
\includegraphics[width=\columnwidth]{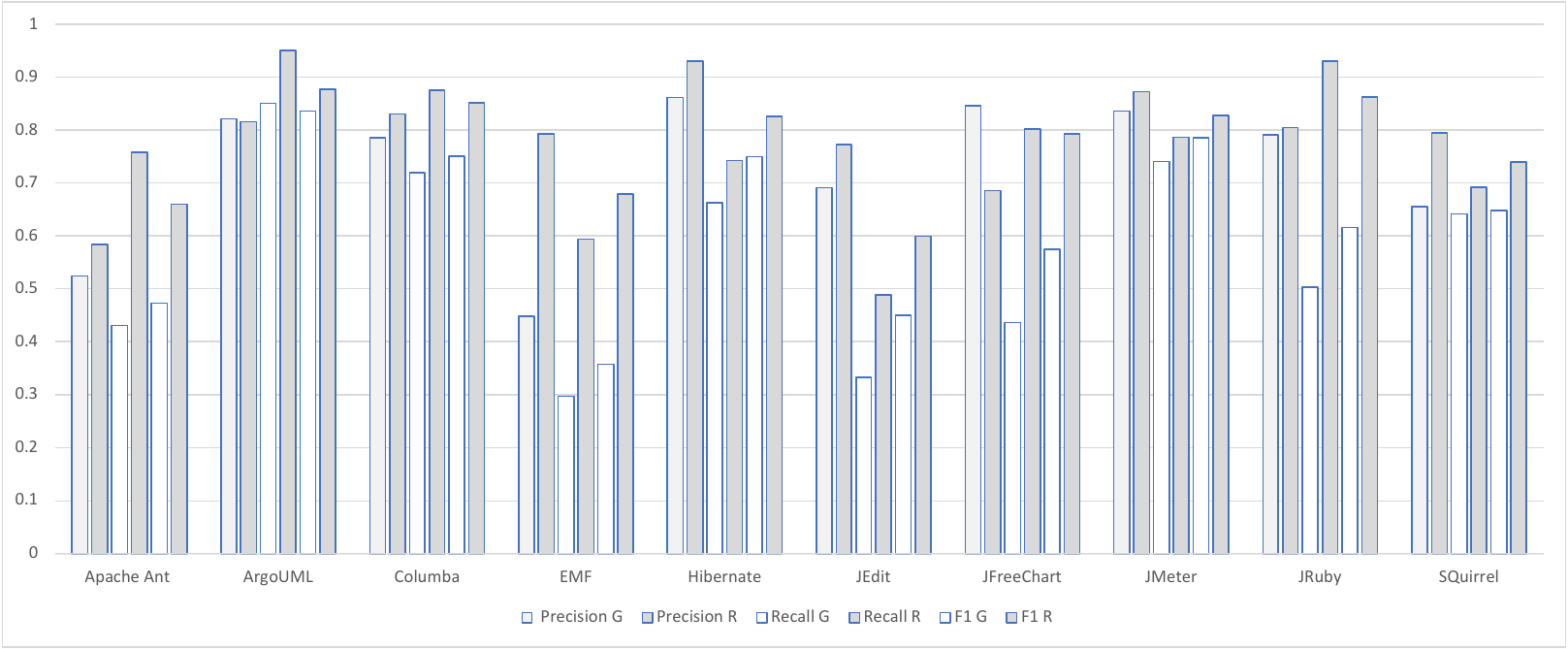}
    \caption{Differences in performance reported by Guo \etal ~of the Ren \etal ~approach~\cite{Guo2021} and in the original work of Ren \etal~\cite{Ren2019TOSEM}}
    \label{fig:performanceDiff}
\end{figure}

Existing public datasets on software data typically provide their own schema and annotation method for a specific learning task (e.g., detection of code weaknesses). 
The schema is often determined by the goal of the research, and labels are typically generated manually,  with heuristics derived for the  available  data or by statistic analyzers, which are not always accurate~\cite{herzig2013impact}. Differences in schema and annotation may prevent study replication and generalization. 
This is, for instance, the case of datasets  annotated for vulnerability detection, for which 
literature reports several issues. 
Vulnerability datasets often  rely on human-labelled techniques (e.g., commit differential analysis) that are resource-intensive ~\cite{zheng2021d2a}.  
A significant number (more than 60\%) of  vulnerabilities miss any annotation  in practice, for example, due to silent fixes (i.e., developers commit changes to fix vulnerabilities but do not label/report the commits ~\cite{TruongEtAl2022}), which implies that the actual number of vulnerabilities is much more than what it can be.
Automated labelling results in a high percentage of incorrectly labelled vulnerabilities ~\cite{TruongEtAl2022}. 
Finally, the different types of annotation techniques may produce different annotated datasets and  detection results~\cite{croft2022data}. 
Another example are techniques to annotate self-admitted technical debt (SATD), i.e., annotate comments and related code (i.e., technical debt) that has low quality and requires future effort for refactoring~\cite{Russo2022}.  
\begin{figure}[!h]
    \centering
\includegraphics[width=\columnwidth]{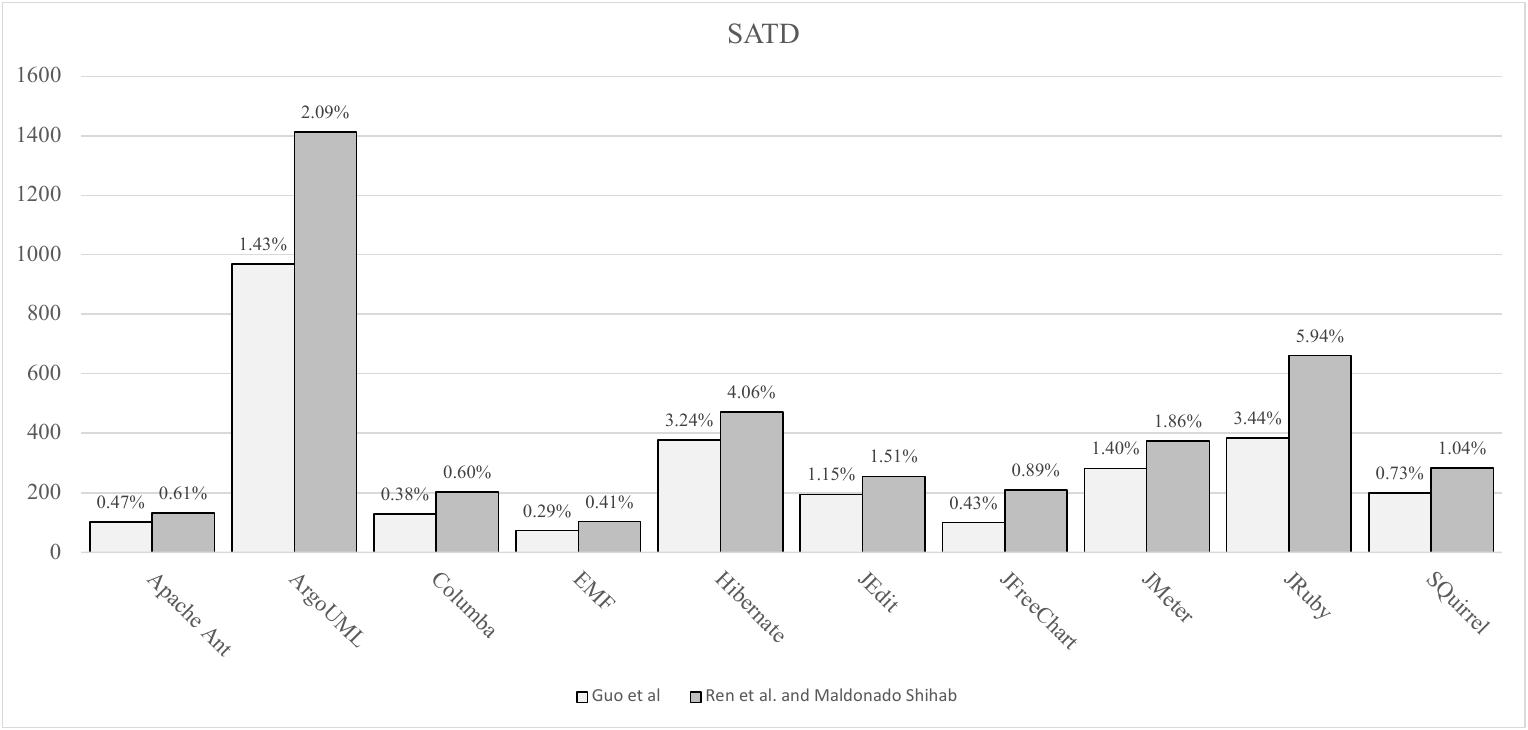}
    \caption{Differences in number and percentage of SATD instances in Ren \etal ~\cite{Ren2019TOSEM} and Guo \etal ~\cite{Guo2021} on the same set of comments~\cite{Maldonado2017}}
    \label{fig:SATDDIff}
\end{figure}
Fig.~\ref{fig:SATDDIff} illustrates the difference in size of the same datasets annotated with the same technique by two different authors Guo \etal~\cite{Guo2021} and Ren \etal~\cite{Ren2019TOSEM}. The figure reports the  number of SATD instances (y-axis) and the percentage of SATD in the respective project. For instance, there is a difference of 2.5\% in SATD instances for JRuby project. 
Differences in datasets' construction and annotation can produce different detection performances, as shown in Fig.~\ref{fig:performanceDiff}.
In this paper, we apply data fusion~\cite{BleiholderNaumann2009} to three existing datasets (WeakSATD~\cite{Russo2022}, Devign~\cite{Zhou2019}, and Big-Vul~\cite{Fan2020}) and build \dataset thereafter, a novel curated dataset of functions and comments (including leading comments, i.e., those comments preceding and related to a function), labelled for vulnerability, technical debt and security concerns.
The dataset provides a unique schema and different annotations for the same instances and the above attributes. 
\dataset~contains about 860k functions and 2.7M of comments from 12 projects. We also propose an approach for the construction and annotation compliant with the schema of \dataset~that enables extension to further projects.
%
The dataset, including the code to create it, is publicly available in the replication package \cite{replication}.
%
\begin{figure*}[b!]
    \centering
    \includegraphics[width=0.9\textwidth]{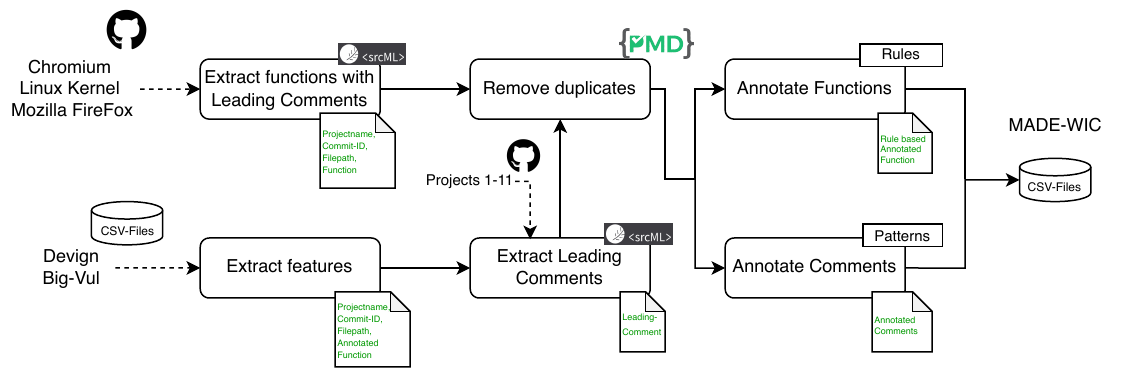}
    \caption{Fusion approach to generate \dataset, extracting the information from existing datasets and open source projects.
    }
    \label{fig:approach}
\end{figure*}

\section{Annotation approaches}
We implemented different techniques with which we labelled the functions of \dataset~either directly or through related comments as described below. 
\subsection{Function Annotations}
\label{sec:annotation}
\textbf{vf:} We indicate with vf the original annotations of Devign and Big-Vul as described in the following.\\
\textit{Devign:} The Devign  approach \cite{Zhou2019} leverages a list of security-related keywords in commits' messages to classify commits as vulnerable, collects the previous version of the  functions changed in the commits and then manually annotates them for vulnerability. \\
\textit{Big-Vul:} The annotation method used in Big-Vul \cite{Fan2020} parses the information in the Common Vulnerability Enumeration (CVE) repository~\cite{CVE} to retrieve the  CVE entries that have reference links to publicly available Git repositories (e.g., GitHub) and their related bugs.  From the related BugID, the fix commits and the previous versions of changed functions are retrieved and labelled vulnerable.\\
\textbf{W:} The W annotation  uses heuristics extracted from the MITRE Common Weakness Enumeration repository (CWE)~\cite{CWE}, as proposed in WeakSATD approach~\cite{Russo2022}. These are a set of rules that implement the descriptions provided in the CWE reports. We used the rules available in the replication package of WeakSATD~\cite{Russo2022}.

\subsection{Function Annotation through Comments}
\textbf{PS:} The PS annotation labels functions as technical debt if they have at least one comment annotated with one of the 64 SATD patterns \cite{Potdar14}. The patterns have been identified by manually inspecting more than 100,000 Java comments.\\
\textbf{MAT:} The MAT  annotation leverages the Matches task Annotation Tags (MAT) heuristics~\cite{Guo2021} to label functions as technical debt if they have at least one comment annotated with one MAT tag \textit{TODO, FIXME, XXX, and HACK}. MAT tags have been identified by exploring the default syntax highlighting of different IDEs.\\
\textbf{SecI:} The SecI annotation first automatically label functions as vulnerable if they have at least one comment annotated with one of the 288 security indicators~\cite{croft2022empirical}. Then, functions and their annotations have been manually reviewed by three of the authors,  achieving a Fleiss' Kappa~\cite{Fleiss1971} score of 0.735 for their agreement. The result is a set of 89 agreed security indicators.
\section{Dataset fusion}
\dataset~fuses and extends two existing datasets Devign \cite{Zhou2019}, and Big-Vul \cite{Fan2020}  by creating a unique schema that transforms the data present in the sources into a common representation. 
The  selected datasets and projects use 
C/C++, provide functions and comments or references to the GitHub projects
from which to extract them, and make their annotation approach
publicly 
available. 
The resulting dataset consists of functions and their comments (internal 
and leading comments).  
Fig.~\ref{fig:approach} illustrates the overall fusion and annotation approach as described in the following sections.
\subsection{Data Extraction:}
To extract the data, we implemented and automated two processes:  1) from the annotated datasets Devign and BigVul and 2) from the open-source repositories starting from the Chromium project used in WeakSATD and extended it to Linux Kernel, Mozilla FireFox.

\textit{From annotated datasets.} From Devign and BigVul, we extracted the features (Project Name, Commit-ID, File Name, Annotation, Function) and used the commit-ID to get the project version. We then used srcML~\cite{CollardEtAl2013}  to obtain any leading comment  of the functions.
Given that retrieving the leading comments is  time-consuming, and Big-Vul is a very large dataset, we choose the ten projects with the largest number of vulnerabilities, accounting for 75\% of the total. The projects are listed in Table~\ref{tab:datasetsDemographics}.
From the table, we can also see that three of the projects (Linux, Chrome, and FFmpeg)  are shared between Devign and BigVul, but the different extraction and annotation techniques make them different  in size and composition.

\textit{From open source public repositories.} 
From open source projects (Chromium, Linux Kernel, Mozilla FireFox), we cloned the repositories at the last commit \footnote{commit hash  57f97b2 for Chromium, e2ca6ba for Linux Kernel, and 4d46db3ff28b for Mozilla Firefox.} and extract with srcML the features (FileName, Annotation, Function, and leading comment, if any). 

Finally, in both processes, we removed duplicates using PMD-CPD~\cite{pmd}, which tokenizes the functions and calculates their similarity based on a threshold of 30 tokens. The threshold balances computational effort with the thoroughness of the inspection and includes 85\% of all functions. Then, we annotated the functions in multiple ways, either directly or through the comments as described in Section~\ref{sec:annotation}. The annotated functions and their comments are then stored in the replication package in three CSV files called OSPR, BigVul, and Devign. 

\begin{table}[!t]
    \centering
    \caption{\dataset~schema.}
    \label{tab:dataset_description}
    \begin{tabular}{p{0.005\columnwidth}p{0.15\columnwidth}p{0.1\columnwidth}p{0.57\columnwidth}} 
    \toprule
    & \textbf{Name} & \textbf{Type} & \textbf{Description} \\
    \midrule 
    & Project-name & string & Project from which the function was extracted \\
    \multirow{4}{*}{\rotatebox[origin=c]{90}{Data}}& Commit-ID & string & Commit hash from which the function was extracted \\ 
    & Filepath & string & Path of the file that contains the function in the selected commit  \\
    & Function & string & Function code \\
    & Leading- \mbox{Comment} & string &  Comment preceding and related to the
function \\
     \hline
     \multirow{6}{*}{\rotatebox[origin=c]{90}{Annotation}}& PS & boolean & Flag indicating PS annotation\\
    & MAT & boolean & Flag indicating MAT annotation\\
    & Big-Vul & boolean & Flag indicating Big-Vul annotation \\
    & Devign & boolean &   Flag indicating Devign annotation\\
    & W & boolean & Flag indicating W annotation\\
    & SecI & boolean &  Flag indicating SecI annotation\\
    \hline
    \end{tabular}
\end{table}

\subsection{Schema}
\label{sec:schema}
Table~\ref{tab:dataset_description} illustrates the  schema  resulting from the fusion: each row describes a column of a CSV file of \dataset. 
The schema contains eleven attributes. 
The annotations' flag indicates the type of annotation as described in Section \ref{sec:annotation}. The annotations Big-Vul and Devign are extracted from the original datasets.
The attributes \textit{Project-name}, \textit{Commit-ID}, and \textit{Filepath} can be used to verify or extend in the future \dataset. Table~\ref{tab:datasetsDemographics} describes the three CSV files compounding \dataset. For each file, we report: (fn) the number of functions, (vf) the number of vulnerable functions as in the original dataset, (W) the number of vulnerable functions annotated with W, and the number of functions annotated as technical debt (with PS and MAT) and as security concerns (SecI). 
In particular, the annotations W, PS, MAT, and SecI are performed on all data in the dataset, see Table~\ref{tab:datasetsDemographics}. Thus, the same schema and annotations allow for a comparison of existing datasets like Devign and BigVul that controls bias related to the construction of the annotated datasets when they are used for vulnerability, technical debt, or security detection.
\subsection{Application Scenarios}
\label{sec:applicationScenario}
Our dataset can be used for various classification tasks that involve functions and/or their comments. 
For instance, it can be used to fine-tune pre-trained deep learning models for text and code (e.g., BERT-based transformers such as CodeBERT~\cite{FengEtAl2020}) for downstream tasks that classify functions as technical debt and/or vulnerable by simply exploiting \dataset~with the PS annotation for technical debt and W for weakness. Studies can also understand the impact of different annotation techniques on the same data by, for instance, comparing \dataset~on vf and W, or the different subsets of \dataset, Devign and BigVul, with their original vf annotations.
\begin{table}[!bt]
    \setlength{\tabcolsep}{2.5pt}
    \centering
    \caption{Functions (including their comments) in \dataset: total (fn),  vulnerability annotated as in original datasets (vf), vulnerability annotated as weaknesses (W), technical debt annotated as (PS) and (MAT), and  security annotated as (SecI).}
    \label{tab:datasetsDemographics}
    \begin{tabular}{lcccccc}
    \hline
        \textbf{Project}& \textbf{fn}& \textbf{vf} & \textbf{ W}  & \textbf{ PS} &  \textbf{MAT} & \textbf{SecI} \\
        \hline
        \multicolumn{7}{c}{\cellcolor{gray!25}\textbf{OSPR}, language: C} \\ \hline
        Chromium & 20,028 &-&5,205&151&508&514\\
        \makecell[l]{Linux Ker.} & 652,726 &-&210,220&5,594&8,444&13,182\\ 
        \makecell[l]{Mozilla Fir.} & 15,380 &-&4,200&116&436&141 \\\cline{1-7}
        Total & 688,134 &-&219,625&5,861&9,388&13,837\\ \hline
        \multicolumn{7}{c}{\cellcolor{gray!25}\textbf{Devign}, language: C} \\ \hline
        FFmpeg & 9,738 &4,961&7,321&1,032&1,366&732 \\ 
        Qemu & 17,544 &7,476&7,785&477&1,378&653\\ \cline{1-7}
        Total & 27,282 &12,437&15,106&1,509&2,744&1,385 \\ \hline
        \multicolumn{7}{c}{\cellcolor{gray!25}\textbf{Big-Vul}, language: C, C++} \\ \hline
        Android & 8,671 &1,267&3,598&53&184&316 \\ 
        Chromium & 77,167 &3,938&10,434&165&569&334\\ 
        FFmpeg & 1,925 &114&1,201&70&99&79\\ 
        \makecell[l]{File(1) comm.} & 294 &49&207&12&7&0\\ 
        ImageMagick & 2,489 &338&1,271&12&11&91\\ 
        Kerberos5 & 832 &140&478&15&31&187\\ 
        Linux Ker.& 46,828 &1,955&18,298&768&886&2,652\\ 
        \makecell[l]{PHP Interp.} &2,669&364&1,580&59&88&173\\ 
        Radare2 & 1,168 &73&722&19&67&19\\ 
        Tcpdump & 778 &210&532&64&110&77\\ \cline{1-7}
        Total & 144,358 &8,448&38,214&1,237&2,042&3,928 \\ \hline
        \textbf{Grand Total} & \textbf{859,774} &\textbf{20,885}&\textbf{272,945}&\textbf{8,607}&\textbf{14,174}&\textbf{19,150}\\\hline
    \end{tabular}
\end{table}
 \dataset~can also be used for the summarization of specific comments (e.g., generating SATD or security-related comments from functions) and masking tasks on functions and/or comments. For instance, researchers can mask PS patterns in comments or W patterns in functions and compare the ability of different transformers to retrieve back the patterns.
\subsection{Data Quality} 
In this section, we leverage the work of Croft \etal~\cite{Croft2023} to assess the quality of \dataset.
The paper provides a set of attributes for high-quality of software vulnerability datasets.   \\
\textit{Accuracy -}
\textit{The degree to which the data has attributes that correctly represent the true value of the intended attribute of a concept or event.} To ensure accuracy, we first decided to use the state-of-the-art annotations. Then, we manually reviewed 31 functions annotated as vulnerable by W (one function per heuristic of the W annotation) and their comments from the OSPR portion.  Each of the 31 functions was independently verified by three authors. In the end, 6 heuristics have been tuned and 10 removed. The authors then agreed on 21 rules and corresponding vulnerable functions. \\
\textit{Uniqueness - }
\textit{The degree to which there is no duplication in records.}\\
To remove duplicated functions, we used PMD-CPD~\cite{pmd} and a 99\% threshold for the Jaccard index of overlapping function code. This process was carried out for each of the datasets, both for individual projects and between different projects. In the end, we removed 127 duplicates.\\
\textit{Consistency - }
\textit{The degree to which data instances have attributes that are free from contradiction and are coherent with other instances.}\\ 
As there is no oracle for the attributes we considered in this work, the different annotation techniques for one attribute (e.g., vulnerability) have reported a portion of functions with positive and negative annotation depending on the technique. For instance, the percentage of functions that are consistently annotated vf and W ranges between 8\% and 45\%, while for technical debt (PS and MAT annotation), it ranges between 0\% and 68\% over individual projects.
\\
\textit{Completeness - }
\textit{The degree to which subject data associated with an entity has values for all expected attributes and related instances.}\\
We provided multiple annotations for all attributes of interest (vulnerability, technical debt, security concerns). Some of the OSPR projects miss the Big-Vul or Devign original annotations and will be matter of future work.
\\
\textit{Currentness - }
\textit{The degree to which data has attributes that are of the right age.}
The OSPR data was extracted and annotated in 2023 with the exception of the Chromium project, whose functions were extracted from the WeakSATD repository~\cite{Russo2022}. The Devign and Big-Vul datasets have been extracted as-is from the original sources~\cite{Zhou2019,Fan2020}. 
\section{Related work and conclusions}\label{sec:relatedWork}
To the best of our knowledge, \dataset~is the first dataset that provides functions, all relevant comments and their annotations for technical debt,  vulnerability and security concerns. 
The three datasets we fused in this study (OSPR, Devign, Big-Vul) have single annotations or different annotation methods.
For instance, WeakSATD~\cite{Russo2022} annotates for SATD files of the Chromium project with PS. In our work, we have reviewed this approach and extracted Chromium functions and  comments from the original source. 
Not all the datasets in the literature could also be fused with our approach. 
For instance, the dataset of Lin \etal~\cite{Lin2020} that can be found in the replication package by Nong \etal~\cite{Nong2022} does not provide the commit-IDs from which the functions were extracted, preventing researchers from retrieving any further data than the ones they published, e.g., leading comment.  
D2A~\cite{zheng2021d2a} is a dataset that leverages six open-source projects at function-level granularity. The annotation was done using a tool-based approach, focusing on the function versions before and after a fixing commit. Due to its large size of 3.7 GB, which includes more than 1.2 million instances, integrating this extensive dataset will be addressed in future work.
\paragraph{Acknowledgments}
Moritz Mock is partially funded by the National Recovery and Resilience Plan (Piano Nazionale di Ripresa e Resilienza, PNRR - DM 117/2023). The work has been funded by project no. EFRE1039 under the 2023 EFRE/FESR program.
\bibliographystyle{ACM-Reference-Format}
\bibliography{ref}


\begin{thebibliography}{23}


\ifx \showCODEN    \undefined \def \showCODEN     #1{\unskip}     \fi
\ifx \showDOI      \undefined \def \showDOI       #1{#1}\fi
\ifx \showISBNx    \undefined \def \showISBNx     #1{\unskip}     \fi
\ifx \showISBNxiii \undefined \def \showISBNxiii  #1{\unskip}     \fi
\ifx \showISSN     \undefined \def \showISSN      #1{\unskip}     \fi
\ifx \showLCCN     \undefined \def \showLCCN      #1{\unskip}     \fi
\ifx \shownote     \undefined \def \shownote      #1{#1}          \fi
\ifx \showarticletitle \undefined \def \showarticletitle #1{#1}   \fi
\ifx \showURL      \undefined \def \showURL       {\relax}        \fi
\providecommand\bibfield[2]{#2}
\providecommand\bibinfo[2]{#2}
\providecommand\natexlab[1]{#1}
\providecommand\showeprint[2][]{arXiv:#2}

\bibitem[Bleiholder and Naumann(2009)]%
        {BleiholderNaumann2009}
\bibfield{author}{\bibinfo{person}{Jens Bleiholder} {and} \bibinfo{person}{Felix Naumann}.} \bibinfo{year}{2009}\natexlab{}.
\newblock \showarticletitle{Data fusion}.
\newblock \bibinfo{journal}{\emph{ACM Comput. Surv.}} \bibinfo{volume}{41}, \bibinfo{number}{1}, Article \bibinfo{articleno}{1} (\bibinfo{date}{jan} \bibinfo{year}{2009}), \bibinfo{numpages}{41}~pages.
\newblock
\showISSN{0360-0300}


\bibitem[Collard et~al\mbox{.}(2013)]%
        {CollardEtAl2013}
\bibfield{author}{\bibinfo{person}{Michael~L. Collard}, \bibinfo{person}{Michael~John Decker}, {and} \bibinfo{person}{Jonathan~I. Maletic}.} \bibinfo{year}{2013}\natexlab{}.
\newblock \showarticletitle{srcML: An Infrastructure for the Exploration, Analysis, and Manipulation of Source Code: A Tool Demonstration} \emph{(\bibinfo{series}{ICSME})}. \bibinfo{pages}{516--519}.
\newblock


\bibitem[Croft et~al\mbox{.}(2022a)]%
        {croft2022empirical}
\bibfield{author}{\bibinfo{person}{Roland Croft} {et~al\mbox{.}}} \bibinfo{year}{2022}\natexlab{a}.
\newblock \showarticletitle{An empirical study of developers’ discussions about security challenges of different programming languages}.
\newblock \bibinfo{journal}{\emph{Empirical Software Engineering}}  \bibinfo{volume}{27} (\bibinfo{year}{2022}), \bibinfo{pages}{1--52}.
\newblock


\bibitem[Croft et~al\mbox{.}(2023)]%
        {Croft2023}
\bibfield{author}{\bibinfo{person}{Roland Croft}, \bibinfo{person}{M.~Ali Babar}, {and} \bibinfo{person}{M.~Mehdi Kholoosi}.} \bibinfo{year}{2023}\natexlab{}.
\newblock \showarticletitle{Data Quality for Software Vulnerability Datasets} \emph{(\bibinfo{series}{ICSE})}. \bibinfo{pages}{121--133}.
\newblock


\bibitem[Croft et~al\mbox{.}(2022b)]%
        {croft2022data}
\bibfield{author}{\bibinfo{person}{Roland Croft}, \bibinfo{person}{Yongzheng Xie}, {and} \bibinfo{person}{Muhammad~Ali Babar}.} \bibinfo{year}{2022}\natexlab{b}.
\newblock \showarticletitle{Data preparation for software vulnerability prediction: A systematic literature review}.
\newblock \bibinfo{journal}{\emph{IEEE Transactions on Software Engineering}} \bibinfo{volume}{49}, \bibinfo{number}{3} (\bibinfo{year}{2022}), \bibinfo{pages}{1044--1063}.
\newblock


\bibitem[Fan et~al\mbox{.}(2020)]%
        {Fan2020}
\bibfield{author}{\bibinfo{person}{Jiahao Fan}, \bibinfo{person}{Yi Li}, \bibinfo{person}{Shaohua Wang}, {and} \bibinfo{person}{Tien~N. Nguyen}.} \bibinfo{year}{2020}\natexlab{}.
\newblock \showarticletitle{A C/C++ Code Vulnerability Dataset with Code Changes and CVE Summaries} \emph{(\bibinfo{series}{MSR '20})}. \bibinfo{publisher}{Association for Computing Machinery}, \bibinfo{address}{New York, NY, USA}, \bibinfo{pages}{508–512}.
\newblock
\showISBNx{9781450375177}


\bibitem[Feng et~al\mbox{.}(2020)]%
        {FengEtAl2020}
\bibfield{author}{\bibinfo{person}{Zhangyin Feng} {et~al\mbox{.}}} \bibinfo{year}{2020}\natexlab{}.
\newblock \showarticletitle{{CodeBERT: A pre-trained model for programming and natural languages}}.
\newblock \bibinfo{journal}{\emph{EMNLP 2020}} (\bibinfo{year}{2020}), \bibinfo{pages}{1536--1547}.
\newblock
\showISBNx{9781952148903}


\bibitem[Fleiss(1971)]%
        {Fleiss1971}
\bibfield{author}{\bibinfo{person}{Joseph~L Fleiss}.} \bibinfo{year}{1971}\natexlab{}.
\newblock \showarticletitle{Measuring nominal scale agreement among many raters.}
\newblock \bibinfo{journal}{\emph{Psychological bulletin}} \bibinfo{volume}{76}, \bibinfo{number}{5} (\bibinfo{year}{1971}), \bibinfo{pages}{378}.
\newblock


\bibitem[Guo et~al\mbox{.}(2021)]%
        {Guo2021}
\bibfield{author}{\bibinfo{person}{Zhaoqiang Guo} {et~al\mbox{.}}} \bibinfo{year}{2021}\natexlab{}.
\newblock \bibinfo{title}{How Far Have We Progressed in Identifying Self-Admitted Technical Debts? A Comprehensive Empirical Study}.
\newblock
\newblock
\showISSN{1049-331X}


\bibitem[Herzig and Zeller(2013)]%
        {herzig2013impact}
\bibfield{author}{\bibinfo{person}{Kim Herzig} {and} \bibinfo{person}{Andreas Zeller}.} \bibinfo{year}{2013}\natexlab{}.
\newblock \showarticletitle{The impact of tangled code changes} \emph{(\bibinfo{series}{MSR})}. IEEE, \bibinfo{pages}{121--130}.
\newblock


\bibitem[Lin et~al\mbox{.}(2020)]%
        {Lin2020}
\bibfield{author}{\bibinfo{person}{Guanjun Lin}, \bibinfo{person}{Wei Xiao}, \bibinfo{person}{Jun Zhang}, {and} \bibinfo{person}{Yang Xiang}.} \bibinfo{year}{2020}\natexlab{}.
\newblock \showarticletitle{Deep Learning-Based Vulnerable Function Detection: A Benchmark}. In \bibinfo{booktitle}{\emph{Information and Communications Security}}, \bibfield{editor}{\bibinfo{person}{Jianying Zhou}, \bibinfo{person}{Xiapu Luo}, \bibinfo{person}{Qingni Shen}, {and} \bibinfo{person}{Zhen Xu}} (Eds.). \bibinfo{publisher}{Springer International Publishing}, \bibinfo{address}{Cham}, \bibinfo{pages}{219--232}.
\newblock
\showISBNx{978-3-030-41579-2}


\bibitem[Maldonado et~al\mbox{.}(2017)]%
        {Maldonado2017}
\bibfield{author}{\bibinfo{person}{Everton da~Silva Maldonado}, \bibinfo{person}{Emad Shihab}, {and} \bibinfo{person}{Nikolaos Tsantalis}.} \bibinfo{year}{2017}\natexlab{}.
\newblock \showarticletitle{Using Natural Language Processing to Automatically Detect Self-Admitted Technical Debt}.
\newblock \bibinfo{journal}{\emph{IEEE Transactions on Software Engineering}} \bibinfo{volume}{43}, \bibinfo{number}{11} (\bibinfo{year}{2017}), \bibinfo{pages}{1044--1062}.
\newblock


\bibitem[MIRTE(2006)]%
        {CWE}
\bibfield{author}{\bibinfo{person}{MIRTE}.} \bibinfo{year}{2006}\natexlab{}.
\newblock \bibinfo{title}{Common Weakness Enumaration}.
\newblock
\newblock
\urldef\tempurl%
\url{https://cwe.mitre.org}
\showURL{%
\tempurl}


\bibitem[MITRE(1999)]%
        {CVE}
\bibfield{author}{\bibinfo{person}{MITRE}.} \bibinfo{year}{1999}\natexlab{}.
\newblock \bibinfo{title}{Common Vulnerabilities and Exposures}.
\newblock
\newblock
\urldef\tempurl%
\url{https://cve.mitre.org}
\showURL{%
\tempurl}


\bibitem[Mock et~al\mbox{.}(2024)]%
        {replication}
\bibfield{author}{\bibinfo{person}{Moritz Mock}, \bibinfo{person}{Jorge Melegati}, {et~al\mbox{.}}} \bibinfo{year}{2024}\natexlab{}.
\newblock \bibinfo{booktitle}{\emph{Replication package}}.
\newblock
\urldef\tempurl%
\url{https://doi.org/10.5281/zenodo.12567874}
\showURL{%
\tempurl}


\bibitem[Nguyen-Truong et~al\mbox{.}(2022)]%
        {TruongEtAl2022}
\bibfield{author}{\bibinfo{person}{Giang Nguyen-Truong} {et~al\mbox{.}}} \bibinfo{year}{2022}\natexlab{}.
\newblock \showarticletitle{HERMES: Using Commit-Issue Linking to Detect Vulnerability-Fixing Commits} \emph{(\bibinfo{series}{SANER})}. \bibinfo{pages}{51--62}.
\newblock


\bibitem[Nong et~al\mbox{.}(2023)]%
        {Nong2022}
\bibfield{author}{\bibinfo{person}{Yu Nong} {et~al\mbox{.}}} \bibinfo{year}{2023}\natexlab{}.
\newblock \showarticletitle{Open Science in Software Engineering: A Study on Deep Learning-Based Vulnerability Detection}.
\newblock \bibinfo{journal}{\emph{IEEE Transactions on Software Engineering}} \bibinfo{volume}{49}, \bibinfo{number}{4} (\bibinfo{year}{2023}), \bibinfo{pages}{1983--2005}.
\newblock


\bibitem[PMD(2024)]%
        {pmd}
PMD \bibinfo{year}{2024}\natexlab{}.
\newblock \bibinfo{booktitle}{\emph{PMD-CPD}}.
\newblock
\urldef\tempurl%
\url{https://pmd.github.io/pmd/pmd_userdocs_cpd.html}
\showURL{%
\tempurl}


\bibitem[Potdar and Shihab(2014)]%
        {Potdar14}
\bibfield{author}{\bibinfo{person}{Aniket Potdar} {and} \bibinfo{person}{Emad Shihab}.} \bibinfo{year}{2014}\natexlab{}.
\newblock \showarticletitle{An Exploratory Study on Self-Admitted Technical Debt} \emph{(\bibinfo{series}{ICSME})}. \bibinfo{pages}{91--100}.
\newblock


\bibitem[Ren et~al\mbox{.}(2019)]%
        {Ren2019TOSEM}
\bibfield{author}{\bibinfo{person}{Xiaoxue Ren} {et~al\mbox{.}}} \bibinfo{year}{2019}\natexlab{}.
\newblock \showarticletitle{Neural Network-Based Detection of Self-Admitted Technical Debt: From Performance to Explainability}.
\newblock \bibinfo{journal}{\emph{ACM Trans. Softw. Eng. Methodol.}} \bibinfo{volume}{28}, \bibinfo{number}{3}, Article \bibinfo{articleno}{15} (\bibinfo{date}{jul} \bibinfo{year}{2019}), \bibinfo{numpages}{45}~pages.
\newblock
\showISSN{1049-331X}


\bibitem[Russo et~al\mbox{.}(2022)]%
        {Russo2022}
\bibfield{author}{\bibinfo{person}{Barbara Russo}, \bibinfo{person}{Matteo Camilli}, {and} \bibinfo{person}{Moritz Mock}.} \bibinfo{year}{2022}\natexlab{}.
\newblock \showarticletitle{WeakSATD: Detecting Weak Self-admitted Technical Debt} \emph{(\bibinfo{series}{MSR})}. \bibinfo{pages}{448--453}.
\newblock


\bibitem[Zheng et~al\mbox{.}(2021)]%
        {zheng2021d2a}
\bibfield{author}{\bibinfo{person}{Yunhui Zheng} {et~al\mbox{.}}} \bibinfo{year}{2021}\natexlab{}.
\newblock \showarticletitle{D2a: A dataset built for ai-based vulnerability detection methods using differential analysis} \emph{(\bibinfo{series}{ICSE-SEIP})}. IEEE, \bibinfo{pages}{111--120}.
\newblock


\bibitem[Zhou et~al\mbox{.}(2019)]%
        {Zhou2019}
\bibfield{author}{\bibinfo{person}{Yaqin Zhou} {et~al\mbox{.}}} \bibinfo{year}{2019}\natexlab{}.
\newblock \bibinfo{booktitle}{\emph{Devign: Effective Vulnerability Identification by Learning Comprehensive Program Semantics via Graph Neural Networks}}.
\newblock \bibinfo{publisher}{Curran Associates}.
\newblock


\end{thebibliography}
\end{document}